\documentclass[aps,prb,superscriptaddress,reprint,showpacs,floatfix]{revtex4-1}
\usepackage[english]{babel}
\usepackage{amsmath,amsthm,amssymb}
\usepackage{amsfonts}
\usepackage{xspace}
\usepackage{bm}
\usepackage{graphicx}
\usepackage[separate-uncertainty = true, exponent-product = \times]{siunitx}
\usepackage{lmodern}
\usepackage{hyperref}
\usepackage{dcolumn}
\usepackage{color}

\definecolor{PRLblue}{RGB}{45,48,146}
\hypersetup{urlcolor=black, colorlinks=true, citecolor=black, linkcolor=black}

\begin{document}

\title{Current-induced spin torque resonance of a magnetic insulator}
\author{Michael Schreier}
\affiliation{\mbox{Walther-Mei\ss ner-Institut, Bayerische Akademie der Wissenschaften,
Garching, Germany}}
\affiliation{Physik-Department, Technische Universit\"at M\"unchen, Garching, Germany}

\author{Takahiro Chiba}
\affiliation{Institute for Materials Research, Tohoku University, Sendai, Japan}

\author{Arthur Niedermayr}
\affiliation{\mbox{Walther-Mei\ss ner-Institut, Bayerische Akademie der Wissenschaften,
Garching, Germany}}
\affiliation{Physik-Department, Technische Universit\"at M\"unchen, Garching, Germany}

\author{Johannes Lotze}
\affiliation{\mbox{Walther-Mei\ss ner-Institut, Bayerische Akademie der Wissenschaften,
Garching, Germany}}
\affiliation{Physik-Department, Technische Universit\"at M\"unchen, Garching, Germany}


\author{Hans Huebl}
\affiliation{\mbox{Walther-Mei\ss ner-Institut, Bayerische Akademie der Wissenschaften,
Garching, Germany}}
\affiliation{Nanosystems Initiative Munich, Munich, Germany}

\author{Stephan Gepr\"ags}
\affiliation{\mbox{Walther-Mei\ss ner-Institut, Bayerische Akademie der Wissenschaften,
Garching, Germany}}

\author{Saburo Takahashi}
\affiliation{Institute for Materials Research, Tohoku University, Sendai, Japan}

\author{Gerrit~E.~W. Bauer}
\affiliation{Institute for Materials Research, Tohoku University, Sendai, Japan}
\affiliation{\mbox{WPI Advanced Institute for Materials Research, Tohoku University,
Sendai, Japan}}
\affiliation{Kavli Institute of NanoScience, Delft University of Technology, Delft, The
Netherlands}

\author{Rudolf Gross}
\affiliation{\mbox{Walther-Mei\ss ner-Institut, Bayerische Akademie der Wissenschaften,
Garching, Germany}}
\affiliation{Physik-Department, Technische Universit\"at M\"unchen, Garching, Germany}
\affiliation{Nanosystems Initiative Munich, Munich, Germany}

\author{Sebastian~T.~B. Goennenwein}
\email{sebastian.goennenwein@wmi.badw-muenchen.de}
\affiliation{\mbox{Walther-Mei\ss ner-Institut, Bayerische Akademie der Wissenschaften,
Garching, Germany}}
\affiliation{Nanosystems Initiative Munich, Munich, Germany}

\date{\today}
\maketitle

\textbf{
Pure spin currents 
transport angular momentum without an associated charge flow. This unique property makes them attractive for spintronics applications, such as torque induced magnetization control in nanodevices~\cite{Pai2012, Liu2012} 
that can be used for sensing, data storage, interconnects and logics. Up to now, however, most spin transfer torque studies focused on metallic ferromagnets~\cite{Demidov2012, Haazen2013}
, while magnetic insulators were largely ignored, in spite of superior magnetic quality factors~\cite{Kajiwara2010, Hamadeh2014}. Here, we report the observation of spin torque-induced magnetization dynamics in a magnetic insulator. Our experiments show that in ultrathin magnetic insulators the spin torque induced magnetization dynamics can be substantially larger than those generated by the Oersted field. This opens new perspectives for the efficient integration of ferro-, ferri-, and antiferromagnetic insulators into electronic devices.
}\\
Magnetic insulators such as yttrium iron garnet (YIG) offer advantages over metallic ferromagnets such as extremely low magnetization damping~\cite{Cherepanov1993}, enabling the long-range transmission of signals via magnetization dynamics, the generation of magnon condensates~\cite{Demokritov2006} or magnonic crystals~\cite{Gulyaev2003}. The absence of mobile charge carriers makes magnetic insulators ideal for applications exploiting the spin degree of freedom. The spin transfer torque and spin pumping~\cite{Tserkovnyak2002} provide the communication channel for exploiting the unique properties of insulating magnetic materials in spintronic devices. However, the relevance of spin transfer torque in magnetic insulators is still controversial~\cite{Miao2014}, although a number of phenomena have been attributed to it~\cite{Kajiwara2010, Chen2013}.\\
The experiments presented here prove that magnetization dynamics in insulators can indeed be actuated by the current-induced spin transfer torque.
We show that we can drive ferromagnetic resonance by a microwave-frequency ($\SI{}{\giga Hz}$) charge current flowing in the Pt layer of a YIG/Pt sample. The ensuing magnetization dynamics is detected by DC spin pumping~\cite{Tserkovnyak2002} and spin Hall magnetoresistance (SMR) rectification~\cite{Nakayama2013, Iguchi2014}. Comparing samples with different YIG film thicknesses, we can discern and quantify the magnetization dynamics driven by the spin transfer torque from that driven by the Oersted field. Even though Pt is not magnetically ordered, we observe a large DC rectification voltage when resonant magnetization dynamics is excited in the YIG. This can naturally be accounted for by the SMR~\cite{Chen2013}. We model the observations with spin diffusion theory and quantum mechanical interface boundary conditions~\cite{Chiba2014, Chiba2014a}, achieving quantitative agreement with the experimental data for samples with different Pt and YIG thicknesses. Our analysis proves that in the sample with the thinnest YIG film, magnetization dynamics are driven by spin transfer torque. This essential observation provides exciting perspectives for spin transfer torque applications with magnetic insulators such as spin transfer torque magnetic random access memory (STT-MRAM) devices  and spin-wave based interconnects.\\

\begin{figure*}%
\includegraphics[width=\textwidth]{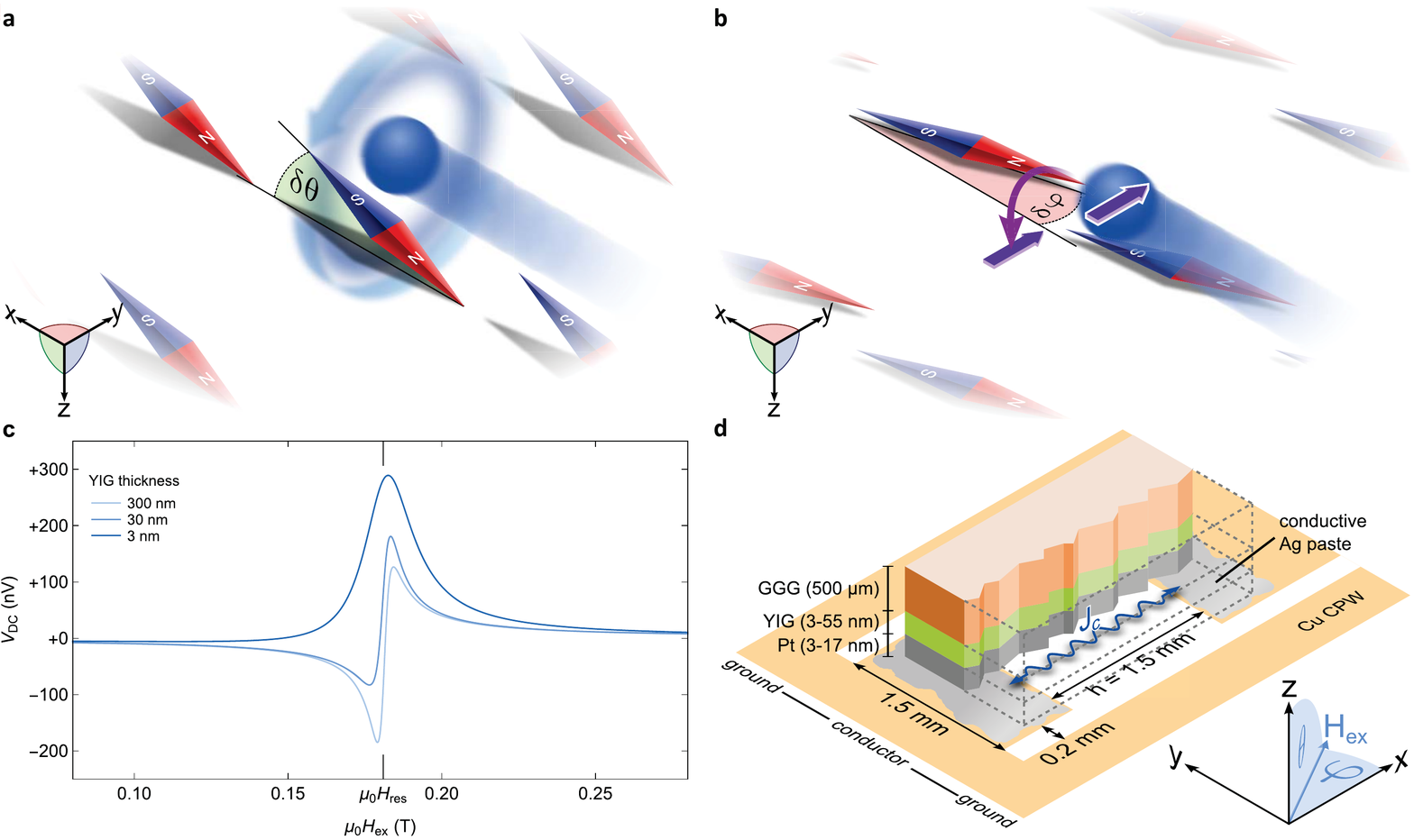}%
\caption{\textbf{Instantaneous effect of spin transfer torque and Oersted fields on the magnetization.} \textbf{a}, The Oersted field generated by a charge current in the $x$ direction, tilts an in-plane magnetization out of the plane by a small angle $\delta\theta$. \textbf{b}, The spin Hall effect generates a pure spin current perpendicular to the original charge current. The torque generated by this spin currents tilts the magnetization of an in-plane magnetized sample by a small angle $\delta\varphi$ in the film plane. For the equilibrium magnetization orientation, the effect of Oersted field and spin transfer torque is reversed. \textbf{c}, The DC voltage generated by the precessing magnetization shows a characteristic dependence on the thickness of the YIG film. While the line is almost purely antisymmetric for thick YIG films, in very thin films the symmetric spin transfer torque contribution dominates \textbf{d}, Our sample is placed across a gap in a coplanar waveguide and contacted with conductive Ag paste.}%
\label{fig:theory}%
\end{figure*}
In conventional magnetic resonance studies with coplanar wave guides the Oersted field generated by the high-fequency current in a metallic wire drives the magnetization precession (Fig.~\ref{fig:theory}a). In the presence of spin-orbit interaction, an AC charge current flowing in a metal is accompanied by a spin current $\pmb{J}_\mathrm{s}$ (Fig.~\ref{fig:theory}b)~\cite{Dyakonov1971, Hirsch1999}. The spin current impinges on the YIG/Pt interface, exerting a (spin transfer) torque on the  magnetization $\pmb{M}$.
A resonantly oscillating spin torque drives a magnetization precession, provided that the spin polarization $\pmb{\sigma}$ of the spin current $\pmb{J}_\mathrm{s}$ and the magnetization $\pmb{M}$ are not collinear.
The magnetization dynamics can be electrically detected via spin pumping (SP)~\cite{Tserkovnyak2002}, 
as well as DC rectification~\cite{Juretschke1960} owing to SMR~\cite{Nakayama2013, Althammer2013}.  
The magnetization dynamics are described by the Landau-Lifshitz-Gilbert equation, including the spin transfer and magnetic field torques:
\begin{equation}
	\partial_t\hat{\pmb{M}}=-\gamma\hat{\pmb{M}}\times\pmb{B}_\mathrm{eff,st}+\alpha_0\hat{\pmb{M}}\times\partial_t\hat{\pmb{M}}+\frac{\gamma\hbar J_{\mathrm{s},z}}{2eM_\mathrm{S}d_\mathrm{F}}\pmb{\sigma}.
\label{eq:LLG}
\end{equation}
Here, $\pmb{B}_\mathrm{eff,st}$ is the sum of external, demagnetization and Oersted fields, $\hat{\pmb{M}}$ is the magnetization unit vector, $\gamma$ is the gyromagnetic ratio, $e$ is the elementary charge, and $M_\mathrm{S}$, $\alpha_0$ and $d_\mathrm{F}$ are the saturation magnetization, intrinsic damping and thickness of the YIG film, respectively. 
Both the Oersted field and the spin transfer torque depend similarly on the magnetization orientation. The two excitation mechanisms are thus difficult to disentangle. Here we study YIG/Pt samples with different layer thicknesses, since the magnitudes of spin transfer torque and Oersted field induced magnetization dynamics differently depend on the thicknesses of the YIG and Pt layers: For a constant charge current density, the Oersted field increases linearly with the thickness $d_\mathrm{N}$ of the Pt layer, but does not depend on $d_\mathrm{F}$. In contrast, the spin transfer can be expressed in terms of an effective (anti-)damping torque field $B_\mathrm{r}$ which is inversely proportional to $d_\mathrm{F}$ and decreases when $d_\mathrm{N}$ exceeds the spin diffusion length~\cite{Chiba2014}. The relative contributions of the driving mechanisms also manifest themselves in the resonance lineshapes that become increasingly symmetric for decreasing $d_\mathrm{F}$ due to the increasing importance of the interface relative to the bulk-induced magnetization torques (Fig.~\ref{fig:theory}c). Both Oersted field driven spin pumping as well as SMR rectification, however, can also contribute a symmetric lineshape~\cite{Chiba2014}. The latter occurs when the phase $\delta$ between the Oersted field and the microwave current is finite, as common for microwaves propagating through magnetic materials. Therefore a quantitative analysis is indispensable for validating the spin torque actuated dynamics.\\

\begin{figure*}[!thb]%
\includegraphics[width=\textwidth]{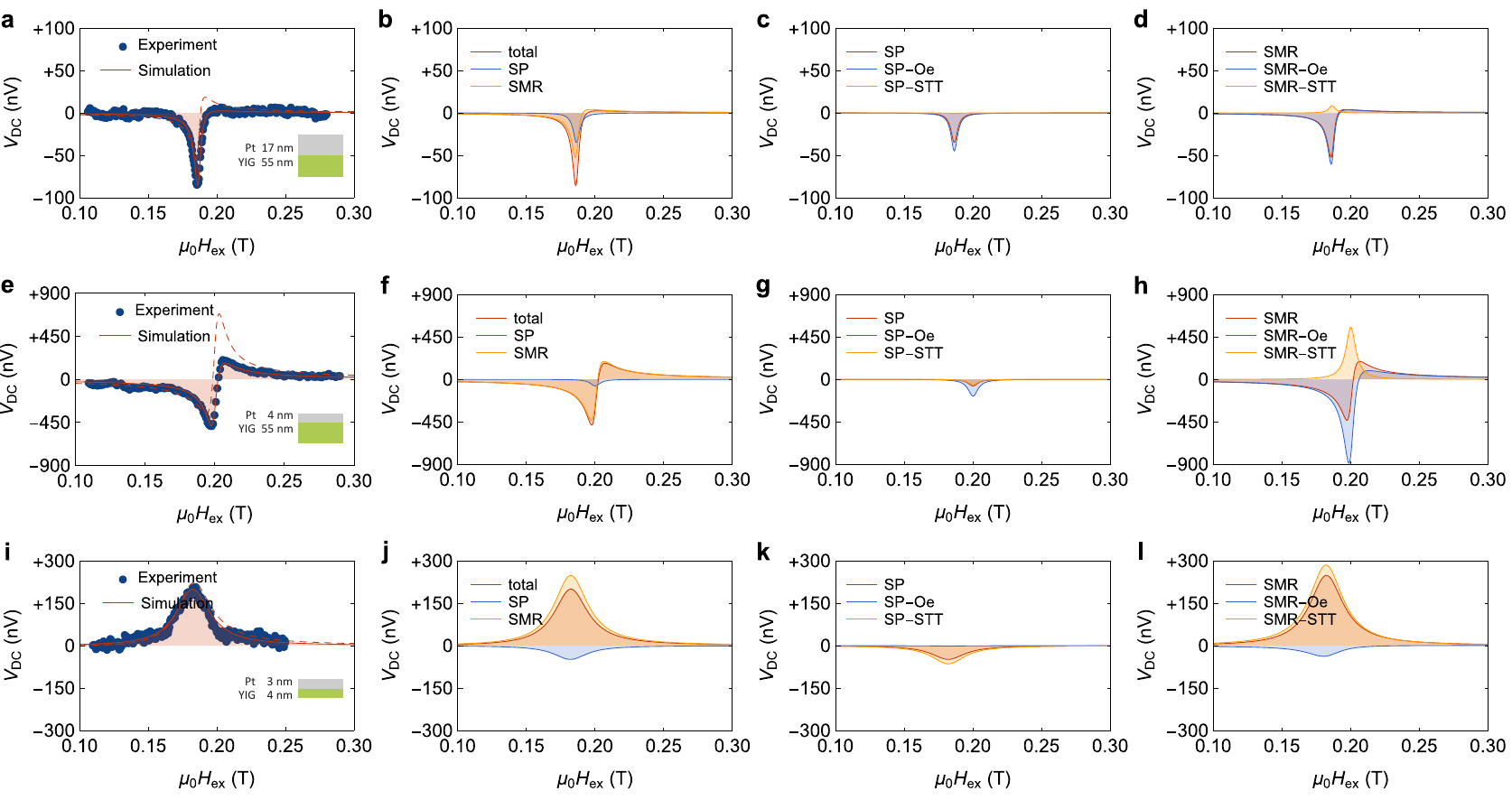}%
\caption{\textbf{Experimental data and simulation of the DC voltages.} \textbf{a}, Measured DC voltage $V_\mathrm{DC}$ of the YIG($\SI{55}{\nano m}$)/Pt($\SI{17}{\nano m}$) sample. The solid red line is calculated from a simulation based on Eq.~\eqref{eq:LLG} for $\delta=-63^\circ$ (contained in $\pmb{B}_\mathrm{eff,st}$) while for the dashed line $\delta=0$ is adopted. The angle between the external magnetic field and the microwave current is $\varphi\cong-35^\circ$ ($\theta=90^\circ$, Fig.~\ref{fig:theory}d). \textbf{b}, Calculated spin pumping (SP) and SMR rectification contributions to the measured DC voltage. \textbf{c, d}, Purely Oersted (Oe) and spin transfer torque (STT) induced contributions to the spin pumping (SP) and SMR rectification components, respectively. \textbf{e-l}, Analogous information for the YIG($\SI{55}{\nano m}$)/Pt($\SI{4}{\nano m}$) and YIG($\SI{4}{\nano m}$)/Pt($\SI{3}{\nano m}$) samples. The relative contributions of the SMR rectification and spin transfer torque components increase with decreasing Pt and YIG thicknesses, respectively.}
\label{fig:results}%
\end{figure*}

Figure~\ref{fig:results} summarizes our results obtained at room temperature for three samples with different film thicknesses: YIG($\SI{55}{\nano m}$)/Pt($\SI{17}{\nano m}$), YIG($\SI{55}{\nano m}$)/Pt($\SI{4}{\nano m}$) and YIG($\SI{4}{\nano m}$)/Pt($\SI{3}{\nano m}$). In panels a, e, and i we plot the DC voltages measured for microwave currents with a frequency $\omega_\mathrm{a}/(2\pi)=\SI{7}{\giga\hertz}$, as a function of the applied magnetic field. A constant offset has been removed from the raw data. For the YIG($\SI{55}{\nano m}$)/Pt($\SI{17}{\nano m}$) sample (panel a) we observe a slightly asymmetric negative voltage peak at the resonance magnetic field. As evident in panel a, the experimental data are well reproduced by our model (for more details, see the supplementary information, Eqns.~(S1) and (S2)). We emphasize that the charge current density $J_\mathrm{c}$ and the intrinsic damping parameter $\alpha_0$ are the only free parameters in the model calculation. Varying impedance matching conditions for the individual samples affects the absolute voltage levels but are naturally accounted for in $J_\mathrm{c}$. Nevertheless, their values are required to be consistent with the applied microwave source power, and the damping parameters extracted from spin pumping experiments on our YIG/Pt samples~\cite{Weiler2013}. The spin transport parameters are taken from earlier, independent experiments: spin diffusion length $\lambda=\SI{1.5}{\nano m}$, spin Hall angle $\theta_\mathrm{SH}=0.11$ and spin mixing conductance $\operatorname{Re}(G^{\uparrow\!\downarrow})=\SI{4e14}{\ohm^{-1}m^{-2}}$ (Refs.~\onlinecite{Althammer2013, Weiler2013}). The phase $\delta=-63^\circ$ is inferred from additional measurements with magnetic fields oriented at a slight angle to the film normal (see supplementary information). The saturation magnetization $M_\mathrm{S}=\SI{118}{\kilo A/m}$ is determined from the magnetic resonance field, and $\rho=\SI{445e-9}{\ohm m}$ from DC resistance measurements. For $J_\mathrm{c}=\SI{0.53e9}{A/m^2}$ and $\alpha_0=0.01$, we obtain good quantitative agreement between model and experiment (Fig.~\ref{fig:results}a). The large intrinsic damping $\alpha_0$ in our samples can be understood in terms of significant two-magnon-scattering induced by roughness in ultrathin ferromagnetic films~\cite{Azevedo2000, Zakeri2007}
, especially when their magnetization lies in the film plane~\cite{McMichael1998}.\\
In panel b we analyze the individual contributions of the SP and SMR components to the DC voltage. Panels c and d show the decomposition of the SP and SMR contributions into Oersted and spin transfer torque (STT) actuated dynamics, respectively. Since the SMR in a $\SI{17}{\nano m}$ thick Pt film is small ($\Delta\rho/\rho\approx\SI{2e-4}{}$)~\cite{Nakayama2013, Althammer2013} spin pumping significantly contributes to the measured voltage signal. The spin pumping signal itself can be attributed almost exclusively to the excitation by the microwave Oersted field (panel c). The same holds for the SMR rectification signal (panel d) as the effective (anti-)damping torque (STT) $\propto d_\mathrm{F}^{-1}$ is much smaller than the Oersted field for thick YIG films. 

In order to separate the Oersted and effective (anti-)damping field (STT) contributions in panel c and d (and analogously in panels g, h, k, l) we set the respective other field to zero. As the spin pumping voltage originates from a nontrivial combination of Oersted field and spin transfer torque induced dynamics, the total spin pumping signal in panel c (g, k) is not just the sum of the individual contributions but depends on $\delta$.\\

We next turn to the YIG($\SI{55}{\nano m}$)/Pt($\SI{4}{\nano m}$) sample. As expected for the reduced Pt thickness, we find a peak and as well as a dip of comparable magnitude in the $V_\mathrm{DC}(H_\mathrm{ex)}$ curve (panel e). This sample exhibits a reduced saturation magnetization and better impedance matching, as reflected by the higher resonance field and absolute voltage level, respectively. Again, the experimental data are well reproduced by the simulation using $M_\mathrm{S}=\SI{89}{\kilo A/m},\ J_\mathrm{c}=\SI{4e9}{A/m^2},\ \rho=\SI{317e-9}{\ohm m},\ \delta=-55^\circ$ and $\alpha_0=0.015$. Separating the contributions from SP and SMR to the measured $V_\mathrm{DC}$ (panel f) reveals a significantly increased SMR rectification contribution as compared to the YIG($\SI{55}{\nano m}$)/Pt($\SI{17}{\nano m}$) sample. This agrees with the expected increase of the SMR magnitude, which has a maximum for a Pt thickness of roughly twice the spin diffusion length~\cite{Althammer2013} $d_\mathrm{N}\approx2\lambda=\SI{3}{\nano m}$. The smaller $d_\mathrm{N}$ enhances the magnitude of the effective (anti-)damping torque. However, since the YIG layer is still comparably thick, the Oersted field contribution to both the SP (panel g) and SMR rectification (panel h) signals still dominates.

Finally, the YIG($\SI{4}{\nano m}$)/Pt($\SI{3}{\nano m}$) sample (panel i) behaves markedly different. Here, we observe a broad positive voltage peak which indicates prominences of STT excitation  (\textit{cf.} Fig.~\ref{fig:theory}c). As the YIG film thickness approaches a single monolayer, the effect of surface roughness on the magnetization damping is increased. This is accurately captured by our simulation for $M_\mathrm{S}=\SI{128}{\kilo A/m},\ J_\mathrm{c}=\SI{1.1e9}{A/m^2},\ \rho=\SI{481e-9}{\ohm m},\ \delta=-78^\circ$ and $\alpha_0=0.04$. The separation into SP and SMR rectification contributions (panel j) shows that the measured voltage signal is now highly symmetric due to the large spin transfer torque component (panel l). The SP component (panel k) is notably smaller than the SMR rectification component.\\
The pronounced dependence of $V_\mathrm{DC}$ on $d_\mathrm{F}$ provides clear evidence for spin transfer torque driven magnetization dynamics, and is faithfully reproduced by our model. 
We thereby identify actively driven, coherent spin transfer torque on a magnetic insulator. 
We intentionally drive resonant magnetization precession, enabling an efficient control of the magnetization dynamics modes in the insulator. Since spin transfer torque is localized at the interface this approach scales advantageously compared to conventional, Oersted field driven magnetization dynamics. Combined with damping reduction by a DC bias current~\cite{Kajiwara2010, Hamadeh2014} the spin transfer torque may be employed to efficiently couple pure magnonic with conventional electronic circuits. Additionally, our results show that AC spin pumping~\cite{Hahn2013a, Wei2014, Weiler2014} in magnetic insulators is reciprocal, as predicted by Onsager symmetry in the linear response regime. With tools such as magnonic mode engineering~\cite{Gulyaev2003} spin transfer torque actuated dynamics may be used even for complex integrated applications.\\

\textbf{Methods}\\
We use an intensity modulated ($f_\mathrm{mod}\cong\SI{10}{\kilo Hz}$) microwave source ($f_\mathrm{MW}=\SI{7}{\giga Hz}$) to feed the samples with an AC charge current. The ensuing DC voltages are detected by a lock-in amplifier. The YIG/Pt samples are integrated into a coplanar waveguide (CPW) structure with a characteristic impedance of $\SI{50}{\ohm}$ and placed onto a $\SI{1.5}{\milli m}$ wide gap in the center conductor. 
The sample dimensions are designed to achieve impedance matching with the microwave circuitry. All YIG/Pt samples cover the entire gap with an effective sample area of about $\SI{1.5}{\milli m}\times\SI{1.5}{\milli m}$. The CPW with the sample attached is placed between the pole shoes of a rotatable magnet.\\
The YIG films were grown in oxygen atmosphere at a pressure of $\SI{25}{\micro\bar}$ on (111) oriented, $\SI{500}{\micro m}$ thick gadolinium gallium garnet (GGG) substrates by laser molecular beam epitaxy. Subsequently the Pt layer was deposited in-situ, without breaking the vacuum, on the YIG thin film by electron beam evaporation~\cite{Althammer2013}. X-ray magnetic circular dichroism (XMCD) measurements~\cite{Geprags2012,*Geprags2013} do not show a proximity effect induced magnetization to below a level of $0.003~\mu_\mathrm{B}/\text{Pt}$.\\
The theory curves in Fig.~\ref{fig:theory}c are calculated using the spin transport parameters given in the main text and using $M_\mathrm{S}=\SI{130}{\kilo A/m}$, $\alpha_0=0.01$, $d_\mathrm{N}=\SI{3}{\nano m}$, $\rho=\SI{300}{\nano\ohm m}$, $J_\mathrm{c}=\SI{2e9}{A/m^2}$, $\varphi=-35^\circ$ and $\delta=0^\circ$.\\

\textbf{Acknowledgements}\\
We thank Sibylle Meyer, Michaela Lammel and Stephan Altmannshofer for the fabrication of YIG/Pt samples. Financial support from the DFG via SPP 1538 ``Spin Caloric Transport'', Project No. GO 944/4-1, BA 2954/1-2, FOM (Stichting voor Fundamenteel Onderzoek der Materie),  the ICC-IMR, the EU-RTN Spinicur, EU-FET grant InSpin 612759 and Grand-in-Aid for Scientific Research (KAKENHI) Nos. 22540346, 25247056, 25220910, and 268063 is gratefully acknowledged.\\

\textbf{Author contributions}\\
M.S. and A.N. performed the measurements, T.C. developed the theory under supervision of G.E.W.B. and S.T., S.G. was responsible for sample fabrication, J.L. and M.S. designed the setup, H.H., R.G. and S.T.B.G. supervised the project. M.S. analyzed the data and prepared the manuscript together with S.T.B.G. and H.H., with input from all authors.


\newpage
\textbf{Supplemental Materials: Current-induced spin torque resonance of a magnetic insulator}
\section{Equations used for the DC voltage simulations}
For the data analysis of the voltage response, we follow Refs.~\onlinecite{Chiba2014, Chiba2014a} detailing the derivation of the measured DC voltage from the Landau-Lifshitz-Gilbert equation [Eq.~(1) in the main text]. Employing the notation of Ref.~\onlinecite{Chiba2014} and converted to SI units, the equation for the DC spin pumping voltage reads
\begin{align}
	  V_\mathrm{SP}=&\frac{h\rho J_\mathrm{r}^\mathrm{P}}{4}\frac{F_\mathrm{S}(B_\mathrm{ex})}{\Delta^2}\cos\varphi\sin2\varphi\nonumber\\
		&\times C\left[C_{+}B_{\rm{ac}}( B_{\rm{ac}}-\alpha B_{\rm{ r}}\cos \delta )\right.\nonumber\\
		&+C_{-}B_{\rm{ r}}( B_{\rm{ r}}+\alpha B_{\rm{ac}}\cos\delta )\nonumber\\
		&+\left.(1 + \alpha^2)C B_{\rm{r}}B_{\rm{ac}} \sin\delta \right]
\label{eq:Vsp}
\end{align}
Here, $h$ is the length of the sample, $\rho$ is its resistivity, $\varphi$ is the in-plane angle between the current and the applied magnetic field $B_\mathrm{ex}=\mu_0H_\mathrm{ex}$, $\delta$ is a phase shift between microwave current and magnetization precession~\cite{Azevedo2011,Bai2013,Iguchi2014} and $\Delta=\alpha\omega_\mathrm{a}/\gamma$ is the magnetic field half width of the resonance determined by the total damping $\alpha$ and the excitation angular frequency $\omega_\mathrm{a}=2\pi f_\mathrm{MW}$. Additionally, we define $C=\tilde{\omega}_a/\sqrt{1+\tilde{\omega}_a^2}$ and $C_\pm=1\pm1/\sqrt{1+\tilde{\omega}_a^2}$ where $\tilde{\omega}_\mathrm{a}=2\omega_\mathrm{a}/(M_\mathrm{S}\gamma\mu_0)$ with the saturation magnetization $M_\mathrm{S}$, gyromagnetic ratio $\gamma$ and vacuum permeability $\mu_0$. Furthermore, we use $J_\mathrm{r}^\mathrm{P}=\hbar\omega_\mathrm{a}/(2ed_\mathrm{N}\rho)\theta_\mathrm{SH}\mathrm{Re}(\eta)$ 
where $e=|e|$ is the elementary charge, $d_\mathrm{N}$ is the thickness of the Pt layer, and $\theta_\mathrm{SH}$ is the spin Hall angle. The parameter $\eta=2\lambda\rho\operatorname{Re}(G^{\uparrow\!\downarrow})\tanh\frac{d_\mathrm{N}}{2\lambda}/[1+2\lambda\rho \operatorname{Re}(G^{\uparrow\!\downarrow})\coth\frac{d_\mathrm{N}}{\lambda}]$ describes the spin diffusion in the Pt layer considering backflow~\cite{Jiao2013}, with the spin diffusion length $\lambda$ and the real part of the spin mixing conductance $G^{\uparrow\!\downarrow}$ (in units of $\SI{}{\ohm^{-1} m^{-2}}$). $B_\mathrm{r}=\hbar/(2eM_\mathrm{S}d_\mathrm{F})\theta_\mathrm{SH}J_\mathrm{c}\mathrm{Re}(\eta)$ is the effective field generated by the (anti-)damping torque for a YIG film thickness of $d_\mathrm{F}$ and a charge current density $J_\mathrm{c}$. $B_\mathrm{ac}=J_\mathrm{c}d_\mathrm{N}\mu_0/2$ is the Oersted field (in close proximity to the Pt film) generated by the microwave current and $F_\mathrm{S}(B_\mathrm{ex})=\Delta^2/[(B_\mathrm{ex}-B_\mathrm{res})^2+\Delta^2]$ describes the resonance with the (Kittel) resonance field $B_\mathrm{res}=-M_\mathrm{S}\mu_0/2+\sqrt{(M_\mathrm{S}\mu_0/2)^2+(\omega_\mathrm{a}/\gamma)^2}$. Finally, the total damping is given by $\alpha=\alpha_0+\gamma\hbar^2/(2e^2M_\mathrm{S}d_\mathrm{F})\operatorname{Re}[G^{\uparrow\!\downarrow}/(1+2\rho\lambda G^{\uparrow\!\downarrow}\coth\frac{d_\mathrm{N}}{\lambda})]$, where $\alpha_0$ is the intrinsic damping of the YIG film.\\
The SMR rectification voltage is given in an analogue manner by
\begin{align}
	V_\mathrm{SMR}=&-\frac{h\Delta\rho_1 J_\mathrm{c}}{4}\frac{F_\mathrm{S}(B_\mathrm{ex})}{\Delta}\cos\varphi\sin2\varphi\nonumber\\
	&\times\left[C(B_\mathrm{r}+\alpha B_\mathrm{ac}\cos\delta)+C_+B_\mathrm{ac}\sin\delta\right.\nonumber\\
	&+\left.B_\mathrm{ac}(C_+\cos\delta-\alpha C\sin\delta)(B_\mathrm{ex}-B_\mathrm{F})/\Delta\right]
\label{eq:Vsmr}
\end{align}
where the resistivity change is given by $\Delta\rho_1=\rho\theta_\mathrm{SH}^2(\lambda/d_\mathrm{N})\mathrm{Re}(\eta)\tanh\frac{d_\mathrm{N}}{2\lambda}$. Note that Eq.~\eqref{eq:Vsp} and Eq.~\eqref{eq:Vsmr} are only valid when $\pmb{B}_\mathrm{ex}$ lies in the film plane. The simulated $V_\mathrm{DC}$ in the main text is the sum $V_\mathrm{SP}+V_\mathrm{SMR}$ where identical parameters are used for $V_\mathrm{SP}$ and $V_\mathrm{SMR}$.\\
\begin{figure*}%
\includegraphics[width=\textwidth]{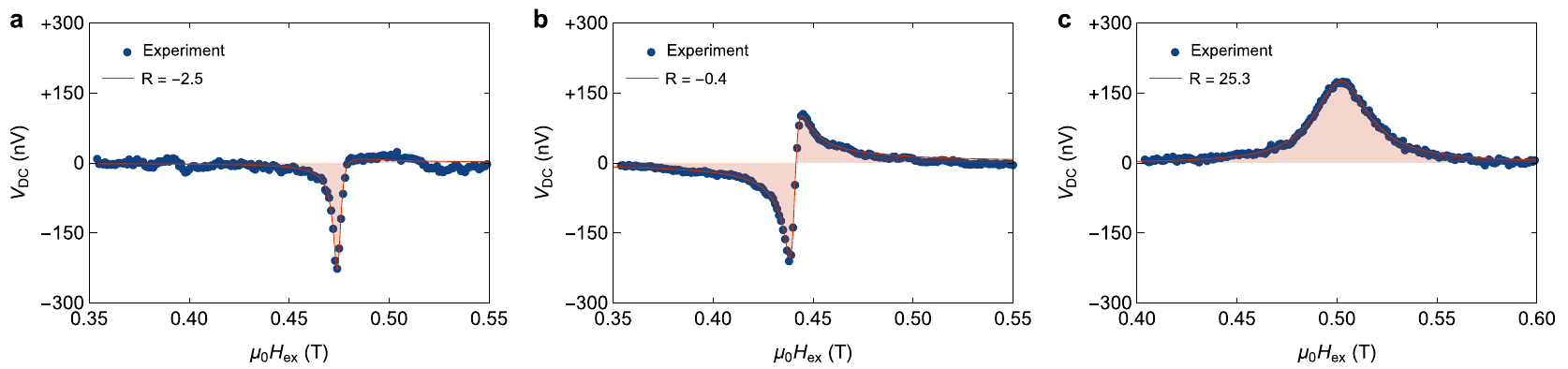}%
\caption{\textbf{Out-of-plane measurements.} Panel a-c show the measured DC voltages for external fields at a slight angle $\theta$ to the film normal on the YIG($\SI{55}{\nano m}$)/Pt($\SI{17}{\nano m}$), YIG($\SI{55}{\nano m}$)/Pt($\SI{4}{\nano m}$) and YIG($\SI{4}{\nano m}$)/Pt($\SI{3}{\nano m}$) sample, respectively.}%
\label{fig:oop}%
\end{figure*}

\section{Out-of-plane measurements and extraction of $\delta$}
As discussed in the main text the measured DC voltages, especially their dependence on the thickness of the YIG and Pt layer, are qualitatively well reproduced by theory. The good agreement is achieved, however, only by taking $\delta\neq0$ in Eqs.~\eqref{eq:Vsp} and \eqref{eq:Vsmr}. Assuming $\delta=0$ in our simulations (dashed lines in Fig.~2 in the main text) yields a notable quantitative disagreement between theory and experiment. For $\delta\neq0$, however, the Oersted field induced SMR rectification voltage [Eq.~\eqref{eq:Vsmr}]can also contribute to the symmetric lineshape. Therefore, quantifying the phases is necessary to unambiguously show that the DC voltages are indeed, at least in part, caused by spin transfer torque rather than by the Oersted field alone. While both $V_\mathrm{SP}$ and $V_\mathrm{SMR}$ vanish for magnetic fields in the plane spanned by film normal and charge current direction~\cite{Bai2013} their magnitude is differently affected by the polar (out-of-plane) angle $\theta$ between the film normal and the magnetization direction~\cite{Chiba2014a}. More specifically, the ratio $R$ of the symmetric to antisymmetric contributions to the lineshape~\cite{Iguchi2014} changes characteristically as a function of $\theta$ for a given $\delta$. A pronounced change in $R$ is observed for $\theta\to0^\circ$ where, however, the DC voltage vanishes. For all samples we thus carried out additional experiments (Fig.~\ref{fig:oop}) with the magnetic field applied at a small angle to the film normal ($\theta\approx5^\circ$, $\varphi=90^\circ$, Fig.~1d in the main text). $\delta$ is then well approximated by extracting $R$ for the in-plane and out-of-plane measurements and adjusting $\delta$ to the unique value yielding both the in-plane and out-of-plane $R$ value. The $R$ value is obtained by fitting the experimental data with a generalized Lorentzian, i.e.
\begin{align}
	V_\mathrm{DC}=&S\frac{{\Delta B}^2}{(B_\mathrm{ex}-B_{res})^2+{\Delta B}^2}\nonumber\\
	+&A\frac{{\Delta B(B_\mathrm{ex}-B_{res})}}{(B_\mathrm{ex}-B_{res})^2+{\Delta B}^2},
\label{eq:R}
\end{align}
where $\Delta B$ is the linewidth, $B_\mathrm{res}$ is the resonance field, $B_\mathrm{ex}$ is the external field and $S$ and $A$ are the amplitudes of the symmetric and antisymmetric contributions to the lineshape. Using $R=S/A$ we extract 
$R_\mathrm{ip}=-3.2$, $R_\mathrm{oop}=-2.5$ [YIG($\SI{55}{\nano m}$)/Pt($\SI{17}{\nano m}$)], $R_\mathrm{ip}=-0.4$, $R_\mathrm{oop}=-0.4$ [YIG($\SI{55}{\nano m}$)/Pt($\SI{4}{\nano m}$)] and $R_\mathrm{ip}=19.4$, $R_\mathrm{oop}=25.3$ [YIG($\SI{4}{\nano m}$)/Pt($\SI{3}{\nano m}$)] from the experimental data. Using identical parametersets we then simulate the in-plane and out-of-plane data using the full theory found in Ref.~\onlinecite{Chiba2014a}. This yields phases of $\delta=-63^\circ$ [YIG($\SI{55}{\nano m}$)/Pt($\SI{17}{\nano m}$)], $\delta=-55^\circ$ [YIG($\SI{55}{\nano m}$)/Pt($\SI{4}{\nano m}$)] and $\delta=-78^\circ$ [YIG($\SI{4}{\nano m}$)/Pt($\SI{3}{\nano m}$)], respectively, to obtain the $R$-values above. Only the $R$ value of the YIG($\SI{4}{\nano m}$)/Pt($\SI{3}{\nano m}$) sample exhibits a notable uncertainty, which translates to a phase error of about $\pm3^\circ$. The indirect procedure employed here is necessary as magnetocrystalline anisotropy other than shape anisotropy is not accounted for in Refs.~\onlinecite{Chiba2014, Chiba2014a} but affects the resonance field in the out of plane measurements. In the simulation we thus assumed that the static magnetization is oriented along the external field direction. Additionally we disregarded all terms associated with the imaginary part of the spin mixing conductance. With out-of-plane magnetic resonance fields of the order of $\SI{450}{\milli T}$ (\textit{cf.} $\mu_0M_\mathrm{S}\lesssim\SI{160}{\milli T}$), $\operatorname{Re}(G^{\uparrow\!\downarrow})\gg\operatorname{Im}(G^{\uparrow\!\downarrow})$ in our YIG/Pt samples~\cite{Althammer2013} and Oersted as well as effective (anti-)damping field both in the film-plane when $\theta\to0^\circ$ these assumptions are expected to introduce only small errors.\\

\bibliography{Bib}
\end{document}